# The Semantic Information Method for Maximum Mutual Information and Maximum Likelihood of Tests, Estimations, and Mixture Models

[1]LU Chenguang

College of Intelligence Engineering and Mathematics, Liaoning Engineering and Technology University, Fuxin, Liaoning, 123000, China

**Abstract**  It is very difficult to solve the Maximum Mutual Information (MMI) or Maximum Likelihood (ML) for all possible Shannon Channels or uncertain rules of choosing hypotheses, so that we have to use iterative methods. According to the Semantic Mutual Information (SMI) and $R(G)$ function proposed by Chenguang Lu (1993) (where $R(G)$ is an extension of information rate distortion function $R(D)$, and $G$ is the lower limit of the SMI), we can obtain a new iterative algorithm of solving the MMI and ML for tests, estimations, and mixture models. The SMI is defined by the average log normalized likelihood. The likelihood function is produced from the truth function and the prior by semantic Bayesian inference. A group of truth functions constitute a semantic channel. Letting the semantic channel and Shannon channel mutually match and iterate, we can obtain the Shannon channel that maximizes the Shannon mutual information and the average log likelihood. This iterative algorithm is called Channels' Matching algorithm or the CM algorithm. The convergence can be intuitively explained and proved by the $R(G)$ function. Several iterative examples for tests, estimations, and mixture models show that the computation of the CM algorithm is simple (which can be demonstrated in excel files). For most random examples, the numbers of iterations for convergence are close to 5. For mixture models, the CM algorithm is similar to the EM algorithm; however, the CM algorithm has better convergence and more potential applications in comparison with the standard EM algorithm.

**Keywords**：Shannon channel, semantic channel, semantic information, likelihood, tests, estimations, mixture model, EM algorithm, machine learning

## 1  Introduction

It is a very important to use the Maximum Mutual Information (MMI) or Maximum Likelihood (ML) as criterion to optimize tests, estimation, predictions, classifications, image compression, and machine learning. The mutual information means saved mean code length and gives higher evaluation to the correct predictions of less probability events. It must be a good criterion when sources are not equiprobable. Yet, the Shannon information theory [1] uses distortion criterion instead of the mutual information criterion to optimize tests and estimations because the optimization needs to change the Shannon channel. Still, without fixing the channel, the mutual information cannot be calculated.

"Maximum Likelihood" [2] appears in many papers, and it has two different meanings. Consider information transfer:

Object $X$—>observed feature $Z \epsilon C$—>hypothesis $Y=f(Z)$

where $X$, $Z$, and $Y$ are discrete random variables and $f(Z)$ is a decision function that ascertains a partition of $C$ and a Shannon channel $P(Y|X)$. The optimal partition will provide the MMI. For a given Shannon channel, there is an ML. For all possible Shannon channels under a certain limit (such as the limit of Gaussian distribution), there is another ML, which is equivalent to the Maximum Average (for different $Y$) Log Likelihood (MALL). This paper focuses on the latter. In the following, "ML" means MALL.

The MMI and ML are so correlative that when we seek an MMI classification, we may need the likelihood method; when we seek an ML estimation (where the probability distribution of all possible hypotheses is uncertain), we could use the concepts of information or entropy. The relationship between the information measure and likelihood has drawn growing attention in recent decades [3, 4, 5].

---

[1] The author's email: lcguang@foxmail.com. Excel files for iterations at http://survivor99.com/lcg/CM-iteration.zip

Can we put likelihood into an information formula to combine Shannon's information theory and the likelihood method more tightly. Akaike notes [3] that the maximum likelihood criterion is equivalent to the minimum Kullback-Leibller (KL) divergence [6] criterion, which is an important discovery. However, the divergence in that context does not mean conveyed information. Although the log relative likelihood ([4], pp. 375-392) and log normalized likelihood [5] used by some researchers are very similar to an information measure; yet they cannot be expressed by the KL information or Shannon Mutual Information (SMI) directly because a sampling distribution is generally different from a likelihood function.

There have been some iterative methods for maximizing the MMI and ML, including the Newton method [7], EM algorithm [8], and minimax method [9]. Still, we want a different iterative method with more efficiency and clearer convergence reasons to use in wider applications.

In a different way, Chenguang Lu, in 1993 [10,11,12], directly defined the semantic information measure (SIM) with the average log normalized likelihood to obtain a generalized KL formula and a generalized mutual information formula. Although he did not use the "likelihood", he used the "predicted probability" which was actually the likelihood. This measure is called the "semantic information measure" because a likelihood function is produced by the truth function of a hypothesis $Y$ and the prior probability distribution of $X$. Lu also proposed the $R(G)$ function, which is an extension of Shannon's rate distortion function $R(D)$ [13], where $G$ is the lower limit of the semantic mutual information or the average log normalized likelihood, and $R$ is the minimum for a given $G$. Now it is found that Lu's information measure and the $R(G)$ function can be used to achieve the MMI and ML more conveniently.

To introduce Lu's method and show its compatibilities with popular likelihood method and popular information theoretical method, it is necessary to clarify the relationship between Lu's truth function and likelihood function. A truth function is equal to a normalized likelihood function with a coefficient so that a truth function has the maximum 1, and takes value from the real interval [0,1]. With the prior probability distribution of a sample or a source, the truth function and likelihood function can be mutually ascertained by Bayesian inference. Whereas a group of truth functions constitute a semantic channel representing the semantic meanings of a group of hypotheses that are understood by the receiver, a Shannon channel represents the rule of using the hypotheses by a sender.

Specifically speaking, the new discovery is that by letting the semantic channel and Shannon channel mutually match and iterate, we can achieve the MMI and ML, not only for tests and estimations, but also for mixture models, for which the EM algorithm [8] is often used. Especially, by the $R(G)$ function, the convergence of the new algorithm can intuitively be explained and proved. We call this algorithm Channels' Matching algorithm, or the CM algorithm.

We first introduce the semantic channel, semantic information measure, and $R(G)$ function in a new way that is as compatible with the likelihood method as possible. Then we discuss how the CM algorithm is applied to tests, estimations, and mixture models with some examples. Finally, we compare the CM algorithm with the EM algorithm to show the advantages and significance of the CM algorithm.

## 2  Semantic Shannel and Semantic Bayesian Inference

A semantic channel consists of a group of truth functions, and affects and is affected by a Shannon channel. First, we discuss the Shannon channel.

### 2.1 The Shannon Channel and the Transition Probability Function

Let $X$ be a discrete random variable representing a fact with alphabet $A=\{x_1, x_2, ..., x_m\}$, let $Y$ be a discrete random variable representing a message with alphabet $B=\{y_1, y_2, ..., y_n\}$, and let $Z$ be a discrete random variable representing a observed condition with alphabet $C=\{z_1, z_2, ..., z_w\}$. A message sender chooses $Y$ to predict $X$ according to $Z$. For example, in weather forecasts, $X$ is a rainfall, $Y$ is a forecast such as "There will be light to moderate rain tomorrow", and $Z$ is a set of meteorological data. In medical tests, $X$ is an infected or uninfected person, $Y$ is positive or negative (testing result), and $Z$ is a laboratory datum or a set of laboratory data.

We use $P(X)$ to denote the probability distribution of $X$ and call $P(X)$ a source, and we use $P(Y)$ to denote the probability distribution of $Y$ and call $P(Y)$ a destination. We call $P(y_j|X)$ with certain $y_j$ and variable $X$ a transition probability function from $X$ to $y_j$. Then a Shannon's channel is composed of a group of transition probability functions [1]:

$$P(Y|X) \Leftrightarrow \begin{bmatrix} P(y_1|x_1) & P(y_1|x_2) & ... & P(y_1|x_m) \\ P(y_2|x_1) & P(y_2|x_2) & ... & P(y_2|x_m) \\ ... & ... & ... & ... \\ P(y_n|x_1) & P(y_n|x_2) & ... & P(y_n|x_m) \end{bmatrix} \Leftrightarrow \begin{bmatrix} P(y_j|X) \\ P(y_j|X) \\ ... \\ P(y_n|X) \end{bmatrix} \quad (2.1)$$

Here a bidirectional arrow means equivalence. The transition probability function has two properties:

1) $P(y_j|X)$ is different from the conditional probability function $P(Y|x_i)$ or $P(X|y_j)$ in that whereas the latter is normalized, the former is not. In general, $\sum_i P(y_j|e_i) \neq 1$.

2) $P(y_j|X)$ can be used to make Bayesian inference to get the posterior probability distribution $P(X|y_j)$ of $X$. To use it by a coefficient, the two inferences are equivalent, i. e.

$$\frac{P(X)kP(y_j|X)}{\sum_i P(x_i)kP(y_j|x_i)} = \frac{P(X)P(y_j|X)}{\sum_i P(x_i)P(y_j|x_i)} = P(X|y_j) \quad (2.2)$$

## 2.2 Semantic Channel and Semantic Bayesian Inference

In terms of hypothesis-testing, $X$ is a sample or a piece of evidence and $Y$ is a hypothesis or a prediction. We need a sample sequence or a sampling distribution $P(X|.)$ to test a hypothesis to see how accurate it is.

Let $\Theta$ be a random variable for a predictive model, and let $\theta_j$ be a value taken by $\Theta$ when $Y=y_j$. The semantic meaning of a predicate $y_j(X)$ is defined by $\theta_j$ or its (fuzzy) truth function $T(\theta_j|X)\in[0,1]$. Because $T(\theta_j|X)$ is constructed with some parameters, we may also treat $\theta_j$ as a set of model parameters (not one of parameters as in the popular method). We can also state that $T(\theta_j|X)$ is defined by a normalized likelihood, i. e., $T(\theta_j|X)=k P(\theta_j|X)/P(\theta_j) = k P(X|\theta_j)/P(X)$, where $k$ is a coefficient that makes the maximum of $T(\theta_j|X)$ be 1. If $T(\theta_j|X)\in\{0,1\}$, $T(\theta_j|X)$ will be the feature function of a set, whose every element makes $y_j$ true. Therefore, $\theta_j$ can also be regarded as a fuzzy set, and $T(\theta_j|X)$ can be considered as a membership function of a fuzzy set defined by Zadeh [14].

In contrast to the popular likelihood method, the above method uses sub-models $\theta_1, \theta_2, ..., \theta_n$ instead of one model $\theta$ or $\Theta$, where a sub-model $\theta_j$ is separated from a likelihood function $P(X|\theta)$ and defined by a truth function $T(\theta_j|X)$. The $P(X|\theta_j)$ here is equivalent to $P(X|y_j, \theta)$ in the popular likelihood method. A sample used to test $y_j$ is also a sub-sample or conditional sample. These changes will make the new method more flexible and more compatible with the Shannon information theory.

When $X=x_i$, $y_j(X)$ become $y_j(x_i)$, which is a proposition with truth value $T(\theta_j|x_i)$. Then there is a semantic channel:

$$T(\theta|X) \Leftrightarrow \begin{bmatrix} T(\theta_1|x_1) & T(\theta_1|x_2) & ... & T(\theta_1|x_m) \\ T(\theta_2|x_1) & T(\theta_2|x_2) & ... & T(\theta_2|x_m) \\ ... & ... & ... & ... \\ T(\theta_n|x_1) & T(\theta_n|x_2) & ... & T(\theta_n|x_m) \end{bmatrix} \Leftrightarrow \begin{bmatrix} T(\theta_1|X) \\ T(\theta_2|X) \\ ... \\ T(\theta_n|X) \end{bmatrix} \quad (2.3)$$

The truth function is also not normalized, and its maximum is 1. Similar to $P(y_j|X)$ and $P(\theta_j|X)$, $T(\theta_j|X)$ can also be used for Bayesian inference, i. e. semantic Bayesian inference (or set-Bayesian inference [10]), to produce likelihood function:

$$P(X|\theta_j) = P(X)T(\theta_j|X)/T(\theta_j)$$
$$T(\theta_j) = \sum_i P(x_i)T(\theta_j|x_i) \quad (2.4)$$

where $T(\theta_j)$ is called the logical probability of $y_j$. We may also write $T(\theta_j)$ as $T(y_j)$. If $T(\theta_j|X) \propto P(y_j|X)$, then semantic Bayesian inference is equivalent to Bayesian inference.

Note that $T(\theta_j)$ is the logical probability of $y_j$, whereas $P(y_j)$ is the probability of choosing $y_j$. The two probabilities are very different. $T(\Theta)$ is also not normalized, and generally there is $T(\theta_1)+T(\theta_2)…+ T(\theta_n)>1$. Consider hypotheses $y_1$="There will be light rain", $y_2$="There will be moderate rain", and $y_3$="There will be light to moderate rain". According to their semantic meanings, $T(\theta_3) \approx T(\theta_1)+T(\theta_2)$; however, we may have $P(y_3)<P(y_1)$. Particularly, when $y_j$ is a tautology, $T(\theta_j)=1$; but $P(y_j)$ is almost 0.

The $P(X|\theta_j)$ is a likelihood function and is also different from $P(X|y_j)$ which is a sampling distribution under the observed condition $Z \epsilon C_j$. Hence $P(X|y_j)=P(X|Z \epsilon C_j)$, which is denoted by $P(Z|C_j)$ for simplicity. Because larger samples are considered in this paper, we do not use a sample sequence in time, but a sampling distribution on the set $A$.

A semantic channel is always supported by a Shannon channel. For weather forecasts, the transition probability function $P(y_j|X)$ indicates the rule of choosing a forecast $y_j$. The rules used by different forecasters may be different and have more or fewer mistakes. Whereas, $T(\theta_j|X)$ indicates the semantic meaning of $y_j$ that is understood by the audience. The semantic meaning is generally publicly defined and may also come from (or be affected by) the past rule of choosing $y_j$. To different people, the semantic meaning should be similar.

## 2.3 Is the Likelihood Function or Truth Function Provided by the GPS's Positioning?

Let us consider the semantic meaning of the small circle (or arrow) on the map on a GPS device. The circle tells where the position of the device is. A clock, a balance, or a thermometer is similar to a GPS device in that their actions may be abstracted as $y_j$="$X \approx x_j$", $j=1, 2, …, n$. The $Y$ with such a meaning may be called an unbiased estimate, and its transition probability functions $P(y_j|X)=P(C_j|X)$, $j=1, 2, …, n$, constitute a Shannon channel. Its semantic channel may be expressed by

$$T(\theta_j|X)=\exp[-|X-x_j|^2/(2d^2)], j=1, 2, …, n \qquad (2.5)$$

where $d$ is the standard deviation.

Consider a particular environment (shown in Figure 1) where a GPS device is used in a car.

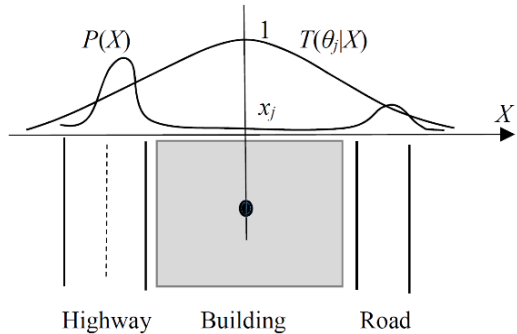

**Figure 1.** An illustration of a GPS's positioning. When the prior probability $P(X)$ is uneven and variable, using a truth function to make a Bayesian inference will be better than using a likelihood function to predict directly

The positioning circle is at a building on the map. The left side of the building is a highway and the right side is a road. We must determine the most possible position of the car. If we think that the circle provides a likelihood function, we should infer "The car is most possibly on the building". However, common sense would indicate that this conclusion is wrong. Alternatively, we can understand the semantic meaning of the circle by a transition probability function. Although this ideal appears sensible, the transition probability function is difficult to obtain, especially when the GPS has a systematical deviation. One may posit that we can use a guessed transition probability function and neglect its coefficient. This idea is a good one. In fact, the truth function in Eq. (2.5) is just such a function. With the truth function, we can obtain the likelihood function by semantic Bayesian inference:

$$P(X|\theta_j) = \frac{P(X)\exp[-(X-x_j)^2/(2d^2)]}{\sum_i P(X)\exp[-(X-x_j)^2/(2d^2)]} \quad (2.6)$$

This likelihood function accords with common sense and avoids conclusion "The car is most likely on the building". The denominator is the logical probability, and it resembles a partition function that often appears in probability theories, information theories, and thermodynamics.

This example shows that a semantic channel is simpler than a Shannon channel. Later, it will be shown that in medical tests, the semantic channel is also simpler and more understandable than the Shannon channel.

## 3 Semantic Information Measure and the Optimization of the Semantic Channel

The Maximum Semantic Information (MSI) estimation bellow is essentially maximum normalized likelihood estimation. Although the MSI estimation is compatible with Maximum Likelihood Estimation (MLE) and maximum likelihood ratio estimation, it can be used in cases where the source is changed.

### 3.1 Defining Semantic Information by Truth Function and Normalized Likelihood

In the Shannon information theory, there is only the statistical probability without the logical probability or likelihood (predicted probability). However, Lu defined semantic information by these three types of probabilities at the same time.

The semantic information conveyed by $y_j$ about $x_i$ is defined as [10]:

$$I(x_i;\theta_j) = \log\frac{P(x_i|\theta_j)}{P(x_i)} = \log\frac{T(\theta_j|x_i)}{T(\theta_j)} \quad (3.1)$$

where semantic Bayesian inference is used; it is assumed that prior likelihood is equal to prior probability. For an unbiased estimation, its truth function and semantic information are illustrated in Figure 2.

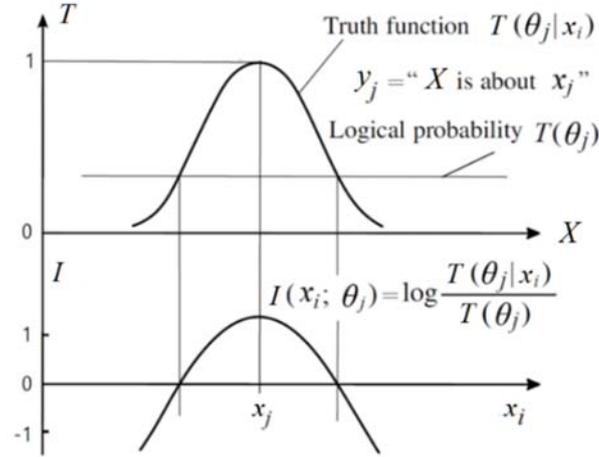

**Figure 2.** Semantic information is defined by log normalized likelihood. The larger the deviation is, the less information there is; the less the logical probability is, the more information there is; lastly, a wrong estimation may convey negative information.

After averaging $I(x_i;\theta_j)$, we obtain semantic (or generalized) KL information:

$$I(X;\theta_j) = \sum_i P(x_i|y_j)\log\frac{P(x_i|\theta_j)}{P(x_i)} = \sum_i P(x_i|y_j)\log\frac{T(\theta_j|x_i)}{T(\theta_j)} \quad (3.2)$$

The statistical probability $P(x_i|y_j)$, $i=1, 2, \ldots$, on the left of "log" above, represents a sampling distribution (note that a sample or sub-sample is also conditional) to test the hypothesis $y_j$ or model $\theta_j$. If $y_j = f(Z|Z \epsilon C_j)$, then $P(X|y_j) = P(X|Z \epsilon C_j)$, which is also denoted by $P(X|C_j)$.

The relationship between the generalized KL information and likelihood is simpler than that between the KL information and likelihood. The KL divergence can be written as the relative entropy $H(P(X)||P(X|y_j))$, which means lost information when we use $P(X)$ as $P(X|y_j)$. The generalized KL information increases with likelihood increasing while the KL information does not. Therefore, the generalized KL information accords with our information concept. It may be negative like log normalized likelihood. It is this property that tells us that wrong predictions or lies may convey negative information.

After averaging $I(X; \theta_j)$, we obtain semantic (or generalized) mutual information:

$$I(X;\Theta) = \sum_j P(y_j) \sum_i P(x_i | y_j) \log \frac{P(x_i|\theta_j)}{P(x_i)}$$
$$= \sum_j \sum_i P(x_i, y_j) \log \frac{T(\theta_j | x_i)}{T(\theta_j)} = H(X) - H(X | \Theta) \tag{3.3}$$
$$H(X | \Theta) = -\sum_j \sum_i P(x_i, y_j) \log P(x_i|\theta_j)$$

Where $H(X)$ is the Shannon entropy of $X$, $\Theta$ is one of a group models $(\theta_1, \theta_2, \ldots, \theta_n)$, $H(X|\Theta)$ is the generalized posterior entropy of $X$. If $P(X)$ is predicted and denoted by $Q(X)$, then there is the relative entropy or KL divergence:

$$H(Q \| P) = \sum_i P(x_i) \log[P(x_i) / Q(x_i)] = H_\Theta(X) - H(X)$$
$$H_\Theta(X) = -\sum_i P(x_i) \log Q(x_i) \tag{3.4}$$

where $H_\Theta(X)$ is the generalized entropy of $X$. It is easy to prove that every generalized entropy above is larger or equal to the corresponding Shannon entropy. They are equal only when a predicted probability distribution totally accords with a statistical probability distribution.

Assume that the size of the sample used to test $y_j$ is $N_j$; the sample points come from independent and identically distributed random variables. Among $N_j$ points, the number of $x_i$ is $N_{ij}$. When $N_j$ is infinite, $P(X|y_j) = N_{ij}/N_j$. Hence there is the following log normalized likelihood:

$$\log \prod_i \left[\frac{P(x_i|\theta_j)}{P(x_i)}\right]^{N_{ji}} = N_j \sum_i P(e_i | y_j) \log \frac{P(x_i|\theta_j)}{P(x_i)} = N_j I(X;\theta_j) \tag{3.5}$$

After averaging the above likelihood for different $y_j$, $j=1, 2, \ldots, n$, we have the average log normalized likelihood (which is equal to the semantic mutual information by $N$ ($N=N_1+N_2+\cdots+N_n$):

$$N \sum_j \frac{N_j}{N} \log \prod_i \left[\frac{P(x_i|\theta_j)}{P(x_i)}\right]^{N_{ji}} = N \sum_j P(y_j) \sum_i P(x_i | y_j) \log \frac{P(x_i|\theta_j)}{P(x_i)} \tag{3.6}$$
$$= NI(X;\Theta) = NH(X) - NH(X | \Theta)$$

It shows that the ML criterion is equivalent to the minimum generalized posterior entropy criterion. When we optimize $\theta_j$ or $\Theta$, $P(X)$ does not change, the MSI criterion is also equivalent to the ML criterion. It is easy to find that when $P(X|\theta_j) = P(X|y_j)$ (for all $j$), the semantic mutual information $I(X;\Theta)$ will be equal to the Shannon mutual information $I(X;Y)$. The latter is the special case of the former, and the former is compatible with the latter.

### 3.2 The Optimization of Semantic Channels

Optimizing a predictive model $\Theta$ is equivalent to optimizing a semantic Channel. For given $y_j$, optimizing $\theta_j$ is equivalent to optimizing $T(\theta_j | X)$ by

$$T^*(\theta_j | X) = \arg\max_{T(\theta_j|X)} I(X;\theta_j) \tag{3.7}$$

$I(X; \theta_j)$ can be written as the difference of two KL divergences:

$$I(X;\theta_j) = \sum_i P(x_i | y_j)\log\frac{P(x_i|y_j)}{P(x_i)} - \sum_i P(x_i | y_j)\log\frac{P(x_i|y_j)}{P(x_i|\theta_j)} \tag{3.8}$$

Because the KL divergence is greater than or equal to 0, when

$$P(X|\theta_j)=P(X|y_j) \tag{3.9}$$

$I(X; \theta_j)$ reaches its maximum and is equal to the KL information $I(X; y_j)$. Let the two sides be divided by $P(X)$; then

$$\frac{T(\theta_j | X)}{T(\theta_j)} = \frac{P(y_j | X)}{P(y_j)} \text{ and } T(\theta_j|X) \propto P(\theta_j|X) \tag{3.10}$$

Set the maximum of $T(\theta_j|X)$ to 1. Then we obtain [16]

$$T^*(\theta_j|X)=P(y_j|X)/P(y_j|x_j^*) \tag{3.11}$$

where $x_j^*$ is the $x_i$ that makes $P(y_j|x_j^*)$ be the maximum of $P(y_j|X)$. Generally it is not easy to get the $P(y_j|X)$; yet for given $P(X|y_j)$ and $P(X)$, it is easy to get $T(\theta_j|X)$ than to get $P(y_j|X)$ since

$$T^*(\theta_j|X)=[P(X|y_j)/P(X)]/[P(x_j^*|y_j)/P(x_j^*)] \tag{3.12}$$

In this equation, $x_j^*$ is such an $x_i$ that makes $P(x_j^*|y_j)/P(x_j^*)$ be the maximum of $P(X|y_j)/P(X)$.

In Eq. (3.3), when $P(Y|X)$ is fixed, we change $T(X|\Theta)$ so that $I(X; \Theta)$ reaches its maximum. This process is called "making the semantic channel match the Shannon channel". When $P(X|\theta_j)=P(X|y_j)$ (for all $j$) or $T(X|\Theta) \propto P(Y|X)$, $I(X; \Theta)$ reaches its maximum, and is equal to the Shannon mutual information $I(X;Y)$. However, conversely, when $T(X|\Theta)$ is fixed, $P(Y|X) \propto T(X|\Theta)$ does not maximize $I(X; \Theta)$. For given $T(X|\Theta)$, there may be other Shannon channels conveying more semantic information. How the Shannon channel matches the semantic channel will be discussed later.

Similar to the Maximum-A-Posteriori (MAP) estimation, the MSI estimation also uses the prior. The difference is that the MAP uses the prior of $Y$, whereas the MSI uses the prior of $X$. The MSI is more compatible with Bayesian inference. The Eq. (3.7) fits parameter estimations, and the Eqs. (3.11) and (3.12) fit non-parameter estimations with larger samples.

## 4 The Matching Function $R(G)$ of Shannon Information and Semantic Information

The $R(G)$ function is an extension of the rate distortion function $R(D)$. It was used to resolve the problem with image compression according to visual discrimination [10-12]. Now it can be used to explain the channels' mutual matching and the CM algorithm.

### 4.1 From the $R(D)$ Function to the $R(G)$ Function

In the $R(D)$ function, $R$ is the information rate, $D$ is the upper limit of the distortion. The $R(D)$ function means that for given $D$, $R=R(D)$ is the minimum of the Shannon mutual information $I(X;Y)$. The (information) rate distortion function with parameter $s$ [15] is

$$\begin{aligned} D(s) &= \sum_i \sum_j d_{ij} P(x_i) P(y_j) \exp(sd_{ij})/\lambda_i \\ R(s) &= sD(s) - \sum_i P(x_i)\ln\lambda_i \end{aligned} \tag{4.1}$$

where $\lambda_i = \sum_j P(y_j)\exp(sd_{ij})$ is the partition function.

Let $d_{ij}$ be replaced with $I_{ij}= I(x_i; y_j)=\log[T(\theta_j|x_i)/T(\theta_j)]= \log[P(x_i|\theta_j)/P(x_i)]$, and let $G$ be the lower limit of the semantic mutual information $I(X; \Theta)$. The $R(G)$ function for a given source $P(X)$ is defined as

$$R(G) = \min_{P(Y|X):I(E;\Theta)\geq G} I(X;Y) \tag{4.2}$$

Following the derivation of $R(D)$, we can obtain [11, 12]

$$G(s) = \sum_i \sum_j I_{ij} P(x_i) P(y_j) 2^{sI_{ij}} / \lambda_i = \sum_i \sum_j I_{ij} P(x_i) P(y_j) m_{ij}^s / \lambda_i$$
$$R(s) = sG(s) - \sum_i P(x_i) \ln \lambda_i \tag{4.3}$$

where $m_{ij}=T(\theta_j|x_i)/T(\theta_j)=P(x_i|\theta_j)/P(x_i)$ is the normalized likelihood; $\lambda_i=\sum_j P(y_i) m_{ij}^s$. We may also use $m_{ij}=P(x_i|\theta_j)$, which results in the same $m_{ij}^s/\lambda_i$. The shape of an $R(G)$ function is a bowl-like curve as shown in Figure 3.

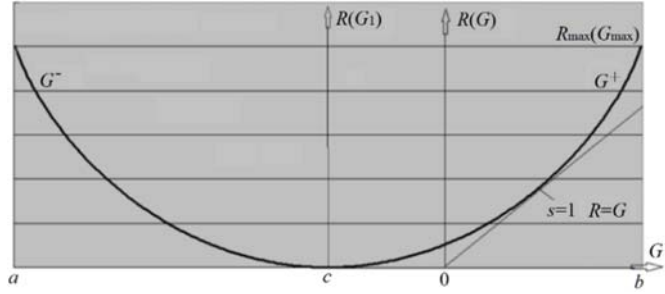

**Figure 3.** The $R(G)$ function of a binary source. As $s=1$, $R=G$, which implies that the semantic channel matches the Shannon channel; $R_{max}(G_{max})$ at the top-right corner means that the Shannon channel matches the semantic channel, hence, both $R$ and $G$ are at their maxima.

The $R(G)$ function is different from the $R(D)$ function. For a given $R$, we have the maximum value $G^+$ and the minimum value $G^-$, which is negative and means that to bring a certain information loss to enemies, we also need certain objective information $R$. When $R=0$, $G$ is negative, which means that if we listen to someone who randomly predicts, the information that we already have will be reduced.

In rate distortion theory, $dR/dD=s$ ($s\leq 0$). It is easy to prove that there is also $dR/dG=s$, where $s$ may be less or greater than 0. The increase of $s$ will reduce the standard deviation of the conditional probability or transition probability distribution, or raise the model's predictive precision.

If $s$ changes from positive $s_1$ to $-s_1$, then $R(-s_1)=R(s_1)$ and $G$ changes from $G^+$ to $G^-$ (see Figure 3).

When $s=1$, $\lambda_i=1$, and $R=G$, which means that the semantic channel matches the Shannon channel and the semantic mutual information is equal to the Shannon mutual information. When $s=0$, $R=0$ and $G(s=0)<0$. In Figure 3, $c=G(s=0)$.

We use the binary source as an example to illustrate the function $R(G)$. Assume

$$I_{ij} = \begin{cases} b>0, & i=j \\ a<0, & i\neq j \end{cases}$$

Following the derivation of $R(D)$ function for a binary source [13, 16], we have

$$R(G) = \frac{b-G}{b-a}\log(b-G) + \frac{G-a}{b-a}\log(G-a) - \log(b-a) + H(X)$$
$$= \frac{h-G_1}{2h}\log(h-G_1) + \frac{h+G_1}{2h}\log(h+G_1) - \log h = H(X) - H(\frac{h-G_1}{2h}) \tag{4.4}$$

where $h=(b-a)/2$, $c=(a+b)/2$, and $G_1=G-c$. The $R(G)$ function is illustrated in Figure 3, where it is assumed that $P(x_0)=P(x_1)=0.5$; $T(\theta_1|x_1)=T(\theta_0|x_0)=1$ and $T(\theta_1|x_0)=T(\theta_0|x_1)=0.2$. Hence $b=0.737$ bits, $a=-1.585$ bits, and $c=-0.424$ bits.

We may define $r=G/R$ as information efficiency, and its maximum is 1 as $s=1$ or $P(X|\theta_j)=P(X|y_j)$, for $j=1, 2, …, n$.

In fact, in the rate distortion theory, if we are allowed to produce a larger error probability, the shape of the function $R(D)$ is also a bowl-like curve. As a result, we can also use a bowl-like $R(D)$ function to optimize the camouflaged messages with greater distortion to puzzle enemies.

## 4.2 Viewing the Maximum Likelihood Ratio Tests from the $R(G)$ Function

For a medical test (see Figure 4), $A=\{x_0, x_1\}$ where $x_0$ means no-infected person and $x_1$ means infected person, and $B=\{y_0, y_1\}$ where $y_0$ means test-negative and $y_1$ means test-positive.

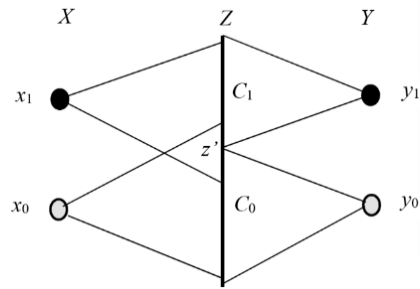

**Figure 4** Illustrating a medical test. It can be abstracted as a 2×2 Shannon nosy channel and the Shannon mutual information changes with partition point $z'$.

In medical tests, the conditional probability in which a test for an infected testee is positive is called sensitivity, and the conditional probability in which a test for an uninfected testee is negative is called specificity [16]. The sensitivity and specificity form a Shannon channel as shown in Table 1.

**Table 1** The sensitivity and specificity of medical tests form a Shannon's channel $P(Y|X)$

| Y | Infected $x_1$ | Uninfected $x_0$ |
|---|---|---|
| Positive $y_1$ | $P(y_1|x_1)$=sensitivity | $P(y_1|x_0)$=1-specificity |
| Negative $y_0$ | $P(y_0|x_1)$=1-sensitivity | $P(y_0|x_0)$=specificity |

If we absolutely believe that a test-positive means being infected and a test-negative means not being infected, then there are truth values $T(y_1|x_1)=T(y_0|x_0)=1$, $T(y_1|e_0)=T(y_0|x_1)=0$. If we use these truth values as the semantic channel, the information will be negatively infinite when one counterexample exists. Thus, we need to consider the confidence levels of $y_j$. Let the confidence level of $y_j$ be denoted by $b$, and let the no-confidence level (i. e. the significance level) be denoted by $b'=1-|b|$. Then the truth function of $y_j$ may be defined as

$$T(\theta_j|X)= b' +bT(y_j|X) \qquad (4.5)$$

Here, $b'$ is actually the truth value of a counterexample.

Assume that the no-confidence level of $y_1$ and $y_0$ are $b_1'$ and $b_0'$ respectively. Assume that the significance level of a medical test is α. The α means that there should be $b_0' \leqslant α$. Table 2 shows the semantic channel for medical tests.

**Table 2.** Two degrees of disbelief of medical tests forms a semantic channel $T(\Theta|X)$

| Y | Infected $x_1$ | Uninfected $x_0$ |
|---|---|---|
| Positive $y_1$ | $T(\theta_1|x_1)=1$ | $T(\theta_1|x_0)=b_1'$ |

| | | |
|---|---|---|
| Negative $y_0$ | $T(\theta_0|x_1)=b_0$' | $T(\theta_0|x_0)=1$ |

According to Eq. (3.11), two optimized no-confidence levels are

$$b_1'^* = P(y_1|x_0)/P(y_1|x_1); \qquad b_0'^* = P(y_0|x_1)/P(y_0|x_0) \tag{4.6}$$

In the medical community, Likelihood Ratio is used to indicate how good a test is [15]. The Eq. (4.6) based on the MSI test is compatible with popular Likelihood Ratio (LR) test. There are

$$LR^+ = P(y_1|x_1)/P(y_1|x_0) = 1/b_1'^*; \quad LR^- = P(y_0|x_0)/P(y_0|x_1) = 1/b_0'^* \tag{4.7}$$

The LR has been used by Thornbury et al for Bayesian inference [17]. Still, it is easier to use the no-confidence level for semantic Bayesian inference. For example[2], $y_1$=HIV-positive, $b_1'^*$=0.0011. If the testees come from ordinary people and $P(x_1)$=0.002, then according to the semantic Bayesian formula (2.4), we have

$$P(x_1|\theta_1) = 0.002/(0.002+0.0011*0.998) = 0.65;$$

If the testees are gay men and $P(x_1)$=0.1, then

$$P(x_1|\theta_1) = 0.1/(0.1+0.0011*0.99) = 0.991.$$

If we predict the infected rate by the likelihood function $P(X|\theta_1)$ directly, after $P(x_1)$ is changed, the likelihood function will be invalid.

Consider the likelihood ratio of tests without a certain partition on $C$:

$$r_L' = \left[\prod_{i=0}^{1}\left(\frac{P(x_i|\theta_1)}{P(x_i|\theta_0)}\right)^{P(x_i|C_1)}\right]^{NP(C_1)} \left[\prod_{i=0}^{1}\left(\frac{P(x_i|\theta_0)}{P(x_i|\theta_1)}\right)^{P(x_i|C_0)}\right]^{NP(C_0)} \tag{4.8}$$

According to Eqs. (3.6) and (4.3), max(log$r_L$)=max(-$NH(X|\Theta)$)-min(-$NH(X|\Theta)$)=$N(G^+ - G^-)$ (see Figure 3). After $R$ and $G^+$ are ascertained, $s$ and $G^-$ are also ascertained. Therefore, the maximum likelihood ratio criterion is equivalent to the maximum likelihood criterion or maximum semantic information criterion.

A binary Shannon channel may be noiseless so that the maximum of $R$ is $R_{max}$=$H(X)$. Yet, for the test shown in Figure 4, noise is inevitable, and hence the $P(Y|X)$ for $R_{max}$<$H(X)$ is not easy to find. As a result, we need an iterative method.

## 5 The CM algorithm for Tests and Estimations

This section will introduce the CM algorithm for tests and estimations, use the $R(G)$ function to explained the iterative convergence, and provide some examples to show iterative processes and speeds.

### 5.1 Explaining Channels' Matching and Iterative Convergence by $R(G)$ Function

**Matching I (Right-step):** The semantic channel matches the Shannon channel.

We keep the Shannon channel $P(Y|X)$ constant, and optimize the semantic channel $T(\Theta|X)$ so that $P(X|\theta_j)=P(X|y_j)$ or $T(\theta_j|X) \propto P(y_j|X)$, and hence $I(X;\Theta)$ reaches its maximum $I(X;Y)$. The aim of the matching I is to let $(R, G)$ move to the point of tangency of curve $R(G)$ and line $R=G$, where $s$=1. See Figure 5 for details.

**Matching II (Left-step):** The Shannon channel matches the semantic channel.

While keeping the semantic channel $T(\Theta|X)$ constant, we change the Shannon channel $P(Y|X)$ to maximize the semantic mutual information $I(X;\Theta)$. In this process, $P(Y|X)$ matches $T(\Theta|X)$. The $R(G)$ function reminds us that $R$ and $G$ can be raised by increasing the parameter $s$. After Matching II, $(G, R)$ locates the top-right corner of a $R(G)$ function curve in Figure 5.

**Matching III (Iteration):** The two channels mutually match and iterate (as shown in Figure 5).

---

[2] https://arxiv.org/abs/1609.07827

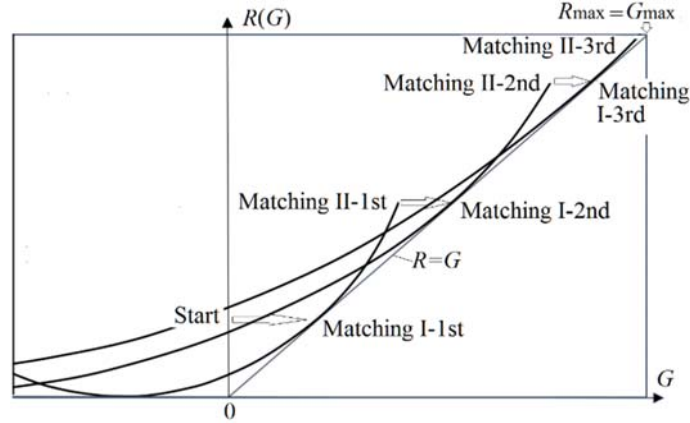

**Figure 5** Illustrating the iterative convergence for tests and estimations. The matching I is for $G=R$; the matching II is to increase $R$ to the top-right corner of a $R(G)$ function; Repeating the matching I and matching II can maximize $R$ and $G$ to obtain $R_{max}$ and $G_{max}$.

The semantic channel and the Shannon channel mutually match alternatively, i. e., iterate, so that $R$ and $G$ reach their maxima at the same time. In terms of likelihood, the predictive model $\Theta$ and the selection rule of model label $Y$ mutually match to achieve maximum average log likelihood. We may say that a semantic channel comes from an old Shannon channel and can be a ladder for a better Shannon channel; then a better Shannon channel can be used to produce a new semantic channel. This process can be repeated.

## 5.2 Three Iterative Examples for Tests and Estimations

It is much simpler to use the CM algorithm for tests and estimations than for mixture models. We do not prove the iterative convergence, which can be intuitively seen by Figure 5. In the following, we use some examples to show the convergence. For the test as shown in Figure 4, optimizing the Shannon channel is equivalent to optimizing the dividing point $z'$. When $Z>z'$, we choose $y_1$; otherwise, we choose $y_0$.

As an example of the test, $Z \in C=\{1, 2, …, 100\}$ and $P(Z|X)$ is a Gaussian distribution function:

$$P(Z|x_1)=K_1\exp[-(Z-c_1)^2/(2d_1^2)], \quad P(Z|x_0)=K_0\exp[-(Z-c_0)^2/(2d_0^2)]$$

where $K_1$ and $K_0$ are normalizing constants. From $P(X)$ and $P(Z|X)$, we can obtain $P(X|Z)$. After setting the starting $z'$, say $z'=50$, as the input of the iteration, we perform the iteration as follows.

**The Right-step (Matching I)**: Calculate the following items in turn.

1) The transition probabilities for the Shannon channel:

$$P(y_0|x_0) = \sum_{z_k=1}^{z'} P(z_k|x_0), \quad P(y_1|x_0) = 1 - P(y_0|x_0)$$

$$P(y_1|x_1) = \sum_{z_k=z'+1}^{100} P(z_k|x_1), \quad P(y_0|x_1) = 1 - P(y_1|x_1)$$

2) The no-confidence levels $b_1'^*$ and $b_0'^*$ according to Eq. (4.6);

3) The logical probabilities $T(\theta_1)=P(x_1)+b_1'^*P(x_0)$ and $T(\theta_0)=P(x_0)+b_0'^*P(x_1)$;

4) $I_{ij}=I(x_i; \theta_j)$ for $i=0, 1$ and $j=0, 1$;

5) The average semantic information $I(X; \theta_1|Z)$ and $I(X; \theta_0|Z)$ for given $Z$ (displaying as two curves):

$$I(X;\theta_j|z_k) = \sum_i P(x_i|z_k)I_{ij}, \quad k=1, 2, …, 100; j=0, 1 \qquad (5.1)$$

**Left-step (Matching II)**: Compare two information function curves $I(X; \theta_1|Z)$ and $I(X; \theta_0|Z)$ over $Z$ to find their cross point. Use this point as new $z'$. If the new $z'$ is the same as the last $z'$ then let $z^*=z'$ and quit the iteration; otherwise go to the Right-step, where $z^*$ is the optimized dividing point.

The following are the reports of three computing examples. Each of Example 2 and Example 3 has two dividing points $z_1'$ and $z_2'$; the iterative principle is the same.

**Iterative Example 1** (for a 2×2 Shannon Channel)

**The Input data**: $P(x_0)=0.8$; $c_0=30$, $c_1=70$; $d_0=15$, $d_1=10$. The start point $z'=50$.

**The iterative process**: Matching II-1 gets $z'=53$; Matching II-2 gets $z'=54$; Matching II-3 get $z^*=54$.

**Comparison**: To see information loss, we get $P(X)=0.72$ bit; $I(X; Z)=0.55$ bit; and $I(X; \Theta)= I(X; Y)=\sum_k P(z_k)I(X; Y|z_k)=0.47$ bit.

**Analysis**: If we use minimum error rate as criterion, the optimal dividing point is 57; yet the above optimal point is $z^*=54$. It is shown that compared with minimum error rate criterion, MSI criterion puts more attention to the correct predictions of small probability events and allow more false positives and less false negatives when $P(x_1)$ is much less than $P(x_0)$.

**Iterative Example 2** (for a 2×3 Shannon channel)

For this channel, if $z_1'<Z\le z_2'$, $Y=y_2=$"The test tells nothing".

**The input data**: $P(x_0)=0.8$; $c_0=30$, $c_1=70$; $d_0=15$, $d_1=10$. The start point $z'_1=50$ and $z_2'=60$.

**The iterative process**: Matching II-1 gets $z_1'=46$ and $z_2'=57$; Matching II-2 gets $z_1'=47$ and $z_2'=59$; Matching II-3 gets $z_1^*=47$ and $z_2^*=59$.

**Comparison and analysis**: $H(X)=0.72$ bit; $I(X; Z)=0.55$ bit; $I(X; Y)=0.52$ bit. Yet in Example 1, $I(X; Y)=0.47$ bit. So, This 2×3 channel can convey more semantic information than the above 2×2 channel. This example shows that we may replace the limit of significance level with the neutral hypothesis $y_2$.

**Iterative Example 3** (for a 3×3 Shannon channel)

This example is to examine a simplified estimation. A pair of good start points and a pair of bad start points are used to check the liability and speed of the iteration.

**The Input data**: $P(x_0)=0.5$, $P(x_1)=0.35$, and $P(x_2)=0.15$; $c_0=20$, $c_1=50$, and $c_2=80$; $d_0=15$, $d_0=15$, $d_1=10$, and $d_2=10$.

**The iterative results**:

1) With the good start points: $z_1'=50$ and $z_2'=60$, the number of iterations is 4; $z_1^*=35$ and $z_2^*=66$.

2) With the bad start points: $z_1'=9$ and $z_2'=20$, the number of iterations is 11; $z_1^*=35$ and $z_2^*=66$ also. Figure 6 shows the information curves over $Z$ before and after the iteration.

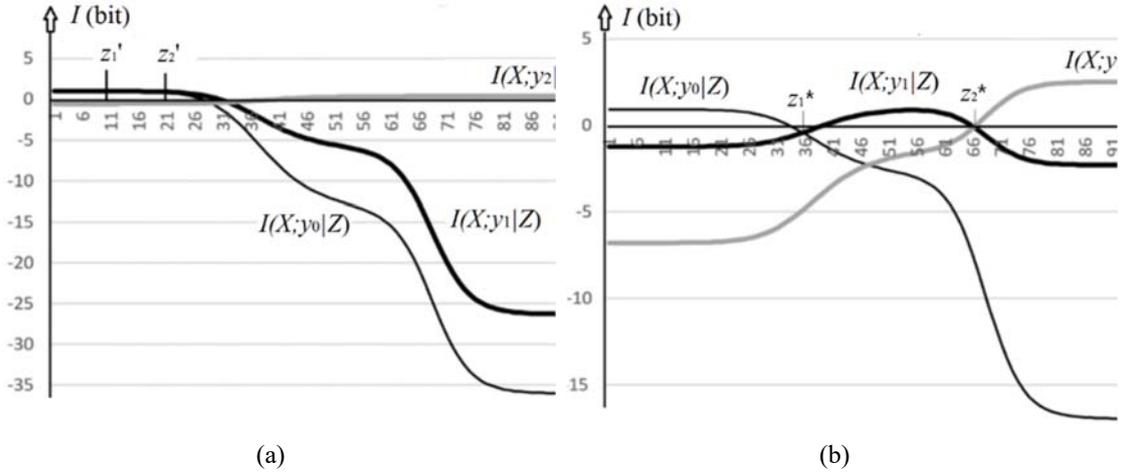

(a)                      (b)

**Figure 6** The iteration with bad start poins. At the beginning of the iteration (a), three information curves have small positive areas. At the end of the iteration (b), three information curves have large positive areas so that $I(X;\Theta)$ reaches its maximum. This figure shows that the iterative convergence is stable even if the start points are very bad.

### 5.3 Analyses and Discussions

The above computing examples show that the convergence of the CM algorithm for tests and estimations is fast and inevitable. Generally, the numbers of iterations for convergence are three to five. In the above examples, $Z$ is one dimensional. If $Z$ is multi-dimensional, calculation must be more complicated; however, the numbers of iterations for convergence should be similar.

The above examples do not use the parameter $s$ of the $R(G)$ function for optimizing the Shannon channel because $z^*$ has contained the information of optimized $s$. The $R(G)$ function with parameter $s$ reminds us that we may use the following equation for fuzzy decision (or classification) function with the consideration of information efficiency $G(s)/R(s)$.

$$P(y_j | z_k) = \frac{P(y_j)[\exp(I(X;\theta_j | z_k))]^s}{\sum_{j'} P(y_{j'})[\exp(I(X;\theta_{j'} | z_k))]^s} , \quad j=1, 2, \ldots, n \qquad (5.2)$$

When $s \to +\infty$, $P(y_j|Z)$ becomes the feature function of $C_j$, as $C$ is optimally partitioned, and tells us the optimal partitioning point $z^*$. Even if $Z$ is multi-dimensional, the above equation is also tenable. This formula can save our energy for searching the boundaries of $C_j$ for all $j$.

We may apply the CM algorithm to general predictions, such as weather forecasts. The application to predictions is similar to the application to estimations. The difference is that the truth functions of predictions may be various. Then we can explain semantic evolution. A Shannon channel indicates a language usage, whereas a semantic channel indicates the comprehension of the audience. The Right-step is to let the comprehension match the usage, and the Left-step is to let the usage (including the observations and discoveries) match the comprehension. The mutual matching and iterating of two channels means that linguistic usage and comprehension mutually match and promote. Natural languages should have been evolving in this way.

If the sampling distribution is discontinuous or irregular so that information curves $I(X;\theta_j|z_k)$ and $I(X;\theta_{j+1}|z_k)$ have more than one cross points, whether may the CM algorithm result in local convergence? This question needs further study.

## 6   The CM algorithm for Mixture Models

This section will solve mixture models by the CM algorithm, prove the iterative convergence with the help of the $R(G)$ function，and use two examples to show the iterative process and convergent speed. Then we compare the CM algorithm with the EM algorithm.

## 6.1 Explaining the Iterative Process by R(G) Function

Assume a sampling distribution $P(X)$ is produced by the conditional probability $P^*(X|Y)$ being some function such as Gaussian distribution. We only know that the number of the mixture components is $n$, without knowing the true $P(Y)$, denoted by $P^*(Y)$. We need to solve $P(Y)$ that is close to $P^*(Y)$ and $P(X|\Theta)$ that is close to $P^*(X|Y)$. The $Z$ and $X$ in tests and estimations now are merged into $X$. A prediction $Y$ is no long correct or wrong, but we require that the predicted probability distribution of $X$, denoted by $Q(X)$, is as close to the sampling distribution $P(X)$ as possible, i. e. $H(Q||P)$ is as small as possible.

Still, we use $P^*(Y|X)$ and $R^*=I^*(X;Y)$ to denote the corresponding Shannon channel and Shannon mutual information. When $Q(X)=P(X)$, there should be $P(X|\Theta)=P^*(X|Y)$, and $G=G^*=R^*$.

Solving maximum likelihood mixture models is different from solving maximum likelihood tests and estimations. For mixture models, when we let the Shannon channel match the semantic channel (in Left-steps), we do not maximize $I(X;\Theta)$, but seek a $P(X|\Theta)$ that accords with $P^*(X|Y)$ as possible (Left-step a in Figure 7 is for this purpose), and a $P(Y)$ that accords with $P^*(Y)$ as possible (Left-step b in Figure 7 is for this purpose). That means we seek a $R$ that is as close to $R^*$ as possible. Meanwhile, $I(X;\Theta)$ may decrease. However, in popular EM algorithms, the objective function, such as $\log P(X^N, Y|\Theta)$, is required to keep increasing without decreasing in both steps.

With CM algorithm, only after the optimal model is obtained, if we need to choose $Y$ according to $X$ (for decision or classification), we may seek the Shannon channel $P(Y|X)$ that conveys the MMI $R_{max}(G_{max})$ (see Left-step c in Figure 7).

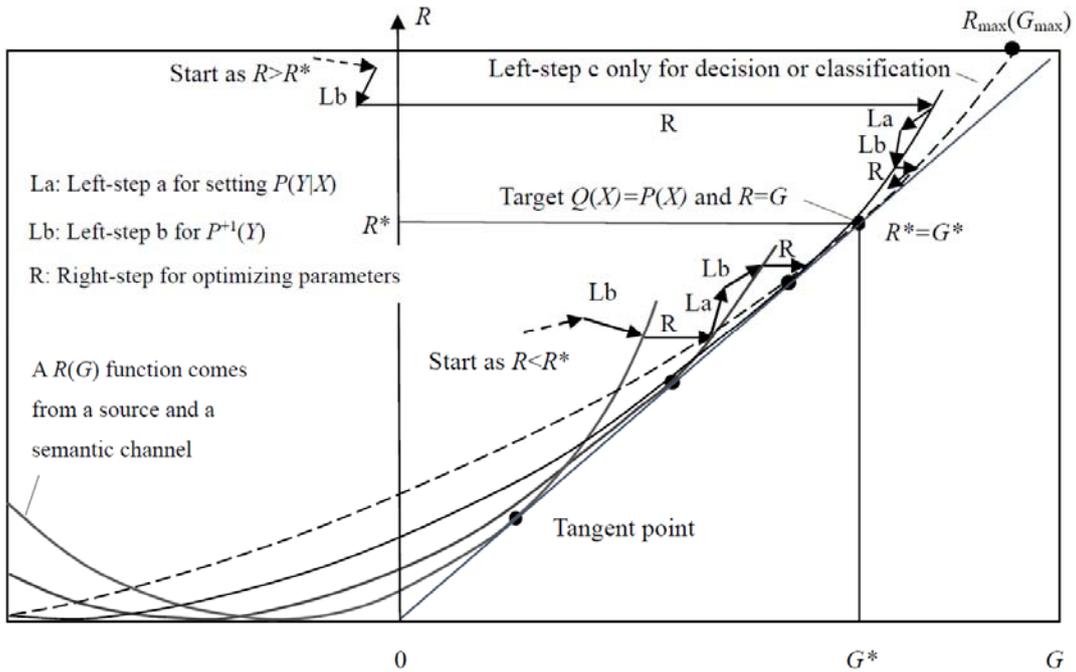

**Figure 7** Illustrating the CM algorithm for mixture models. There are two iterative examples. One is for $R>R^*$ and another is for $R<R^*$. The Left-step a and Left-step b make $R$ close to $R^*$, which means $(G, R)$ is longitudinally improved; whereas the Right-step increases $G$ so that $(G, R)$ approaches line $R=G$, which means $(G, R)$ is horizontally improved.

If we guess that $P(X)$ is produced by $P^*(X|Y)$ with the Gaussian distribution, then likelihood functions are
$$P(X|\theta_j)= k_j \exp[-(X-c_j)^2/(2d_j)^2], j=1,2,\ldots, n$$
Assume $n=2$, then parameters are $c_1, c_2, d_1, d_2$. In the beginning of the iteration, we may set $P(y_1)=P(y_2)=1/2$. We begin iterating from Left-step a.

**Left-step a** Construct Shannon channel by

$$P(y_j | X) = P(y_j)P(X | \theta_j) / Q(X), \quad Q(X) = \sum_j P(y_j)P(X | \theta_j), \quad j=1, 2, ..., n \quad (6.1)$$

This formula has been used in the EM algorithm [8]. It was also used in the derivation process of the $R(D)$ function [16]. Hence the semantic mutual information is

$$I(X;\Theta) = \sum_i \sum_j P(x_i) \frac{P(x_i | \theta_j)}{Q(x_i)} P(y_j) \log \frac{P(x_i | \theta_j)}{P(x_i)} \quad (6.2)$$

**Left-step b** Use the following equation to obtain a new $P(Y)$ repeatedly until the iteration converges.

$$P(y_j) \Leftarrow \sum_i P(x_i)P(y_j | x_i) = \sum_i P(x_i) \frac{P(x_i | \theta_j)}{\sum_k P(y_k)P(x_i | \theta_k)} P(y_j) \quad j=1, 2, ..., n \quad (6.3)$$

The convergent $P(Y)$ is denoted by $P^{+1}(Y)$. This is because $P(Y|X)$ from Eq. (6.1) is an incompetent Shannon channel which makes $\sum_i P(x_i)P(y_j|x_i) \neq P(y_j)$. The above iteration makes $P^{+1}(Y)$ match $P(X)$ and $P(X|\Theta)$ better. This iteration has been used by Byrne for an improved EM algorithm [18].

When $n=2$, we should avoid choosing $c_1$ and $c_2$ so that both are larger or less than the mean of $X$; otherwise $P(y_1)$ or $P(y_2)$ will be 0, and cannot be larger than 0 later. If $n>2$, we may keep any $P(y_j)$ from being too small by stopping the iteration when $P(y_j)<0.1/n$. Once we find a $P(Y)$ that is competent in a Left-step b, we should have found a way for $(G, R)$ to converge to $(G^*, R^*)$. In later step b, we should allow some $P(y_j)$ to be equal to 0.

If $H(Q||P)$ is less than a small number, such as 0.001 bit, then end the iteration; otherwise continue.

**Right-step:** Optimize the parameters in the likelihood function $P(X|\Theta)$ on the right of the log in Eq. (6.2) to maximize $I(X; \Theta)$. Then go to Left-step a.

**About Left-step c** After $P(X|\Theta)$ is optimized, perhaps we need to select $Y$ (making decision or classification) according to $X$. The parameter $s$ in $R(G)$ function (see Eq. (4.3)) reminds us that we may use the following fuzzy decision functions

$$P(y_j | X) = P(y_j)[P(X | \theta_j)]^s / Q(X), \quad Q(X) = \sum_j P(y_j)[P(X | \theta_j)]^s, \quad j=1, 2, ..., n \quad (6.4)$$

When $s \to +\infty$, the fuzzy decision will become crisp decision. Different from Maximum A prior (MAP) estimation, the above decision function still persists in the ML criterion or MSI criterion. The Left-step c in Figure 7 shows that $(G, R)$ moves to $(G_{max}, R_{max})$ with $s$ increasing.

## 6.2 Using Two Examples to Show the Iterative Processes

### 6.2.1 Example 1 for R<R*

In Table 3 there are real parameters that produce the sample distribution $P(X)$ and guessed parameters that are used to produce $Q(X)$. The convergence process from the starting $(G, R)$ to $(G^*, R^*)$ is shown by the iterative locus as $R<R^*$ in Figure 7. The number of iterations is 5 (5 Right-steps and 6 Left-steps). The iterative process is shown in Figure 8. The iterative results are shown in Table 3 and Figure 9.

Table 3 Real and guessed model parameters and iterative results of Example 1 ($R<R^*$)

|  | Real parameters in $P^*(X|Y)$ and $P^*(Y)$ | | | Starting parameters & $P(Y)$; $H(Q||P)$=0.410 bit | | | Parameters after 5 Right-steps & $P(Y)$; $H(Q||P)$=0.00088 bit | | |
|---|---|---|---|---|---|---|---|---|---|
|  | c | d | $P^*(Y)$ | c | d | $P(Y)$ | c | d | $P(Y)$ |
| $y_1$ | 35 | 8 | 0.7 | 30 | 15 | 0.5 | 35.4 | 8.3 | 0.720 |
| $y_2$ | 65 | 12 | 0.3 | 70 | 10 | 0.5 | 66.2 | 11.4 | 0.280 |

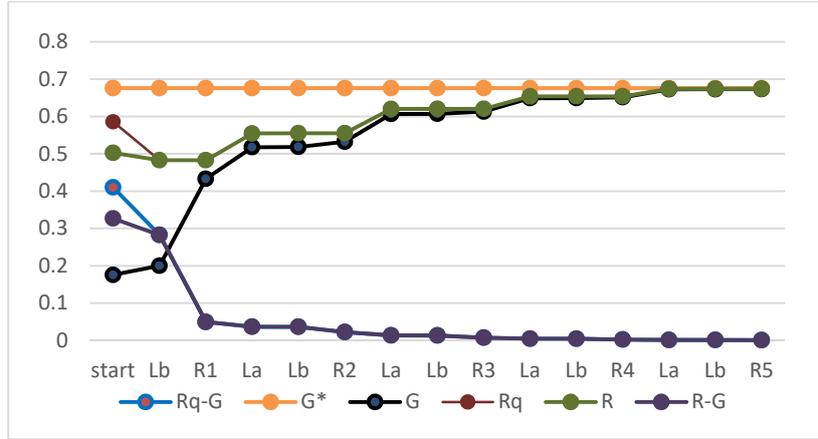

**Figure 8** The iterative process as $R<R^*$. The $Rq$ is $R_Q$ in Eq. (6.4). $H(Q||P)=R_Q-G$ decreases in all steps. $G$ is monotonically increasing. $R$ is also monotonically increasing except in the first Left-step b. $G$ and $R$ gradually approach $G^*=R^*$ so that $H(Q||P)=R_Q-G$ is close to 0.

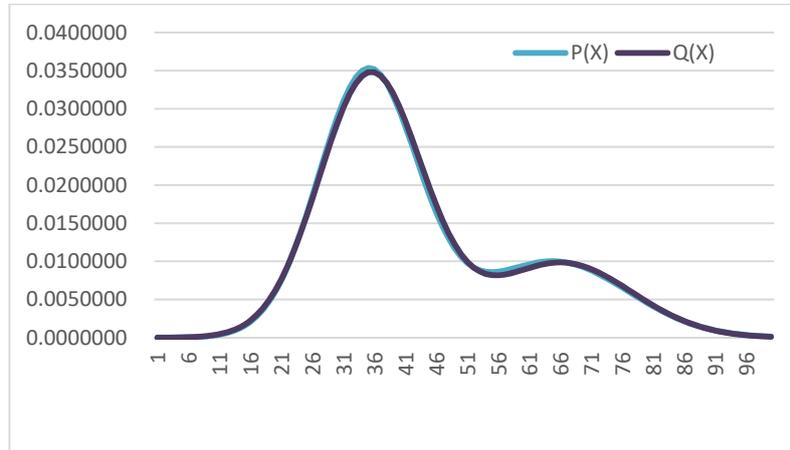

**Figure 9** Comparison of the predicted distribution $Q(X)$ with the sampling distribution $P(X)$ after 5 iterations

**Analyses:** In this iterative process, there are always $R<R^*$ and $G<G^*$. After each step, $R$ and $G$ increase a little bit so that $G$ approaches $G^*$ gradually. This process seams to tell us that each of Right-step, Left-step a, and Left-step b can increase $G$; and hence maximizing $G$ can minimize $H(Q||P)$, which is our goal. Yet, it is wrong. The Left a and Left b do not necessarily increase $G$. There are many counterexamples. Fortunately, the iteration for theses counterexamples still converges. Let us see Example 2 as a counterexample.

### 6.2.2 Example 2 for $R>R^*$

Table 4 shows the parameters and iterative results. The iterative process is shown in Figure 10.

**Table 4 Real and guessed model parameters and iterative results for Example 2 ($R>R^*$)**

|  | Real parameters in $P^*(X|Y)$ & $P^*(Y)$ | | | Starting parameters & $P(Y)$; $H(Q||P)$=0.680 bit | | | Parameters after 5 Right-steps & $P(Y)$; $H(Q||P)$=0.00092 bit | | |
|---|---|---|---|---|---|---|---|---|---|
|  | c | d | $P^*(Y)$ | c | d | $P(Y)$ | c | d | $P(Y)$ |
| $y_1$ | 35 | 8 | 0.1 | 30 | 8 | 0.5 | 38 | 9.3 | 0.134 |
| $y_2$ | 65 | 12 | 0.9 | 70 | 8 | 0.5 | 65.8 | 11.5 | 0.866 |

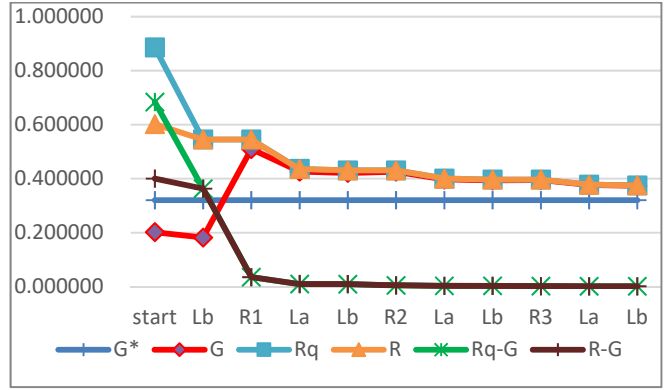

**Figure 10** The iterative process as $R>R^*$. The $Rq$ is $R_Q$ in Eq. (6.4). $H(Q||P)=R_Q-G$ decreases in all steps. $R$ is monotonically decreasing. $G$ increases more or less in all Right-steps and decreases in all Left-steps. $G$ and $R$ gradually approach $G^*=R^*$ so that $H(Q||P)=R_Q-G$ is close to 0.

**Analyses**: $G$ is not monotonically increasing nor monotonically decreasing. It increases in all Right steps and decreases in all Left steps. This example is a challenge to all authors who prove that the standard EM algorithm or a variant EM algorithm converges.

### 6.3 The Proof and Explanation of the Convergence of the CM Algorithm

**Proof** To prove the CM algorithm converges, we need to prove that $H(Q||P)$ is decreasing or no-increasing in every step.

**Consider Right-step.** Assume that the Shannon mutual information conveyed by $Y$ about $Q(X)$ is $R_Q$, and that about $P(X)$ is $R$. Then we have

$$R_Q = I_Q(X;Y) = \sum_i \sum_j P(x_i) \frac{P(x_i|\theta_j)}{Q(x_i)} P(y_j) \log \frac{P(x_i|\theta_j)}{Q(x_i)} \qquad (6.5)$$

$$R = I(X;Y) = \sum_i \sum_j P(x_i) \frac{P(x_i|\theta_j)}{Q(x_i)} P(y_j) \log \frac{P(x_i|\theta_j)}{P(x_i)}$$

$$= \sum_i \sum_j P(x_i) \frac{P(x_i|\theta_j)}{Q(x_i)} P(y_j) \log \frac{P(y_j|x_i)}{P^{+1}(y_j)} = R_Q - H(Y||Y^{+1}) \qquad (6.6)$$

$$H(Y||Y^{+1}) = \sum_j P^{+1}(y_j) \log[P^{+1}(y_j)/P(y_j)]$$

Comparing $G$ in Eq. (6.1) and $R_Q$ in Eq. (6.4), we can find

$$H(Q||P) = R_Q - G = R + H(Y||Y^{+1}) - G \qquad (6.7)$$

In Right-steps, the Shannon channel and $R_Q$ does not change, $G$ is maximized. Therefore $H(Q||P)$ is decreasing and its decrement is equal to the increment of $G$.

**Consider Left-step a.** After this step, $Q(X)$ becomes $Q^{+1}(X)=\sum_j P(y_j)P^{+1}(X|\theta_j)$. Since $Q^{+1}(X)$ is produced by a better likelihood function and the same $P(Y)$, $Q^{+1}(X)$ should be closer to $P(X)$ than $Q(X)$, i. e. $H(Q^{+1}||P) < H(Q||P)$.

**Consider Left-step b**. The iteration for $P^{+1}(Y)$ moves $(G, R)$ to the $R(G)$ function cure ascertained by $P(X)$ and $P(X|\Theta)$. This conclusion can be obtained from the derivation processes of $R(D)$ function [15] and $R(G)$ function [12]. A similar iteration is used for $P(Y|X)$ and $P(Y)$ in deriving the $R(D)$ function. Because $R(G)$ is the minimum $R$ for a given $G$, hence $H(Q||P)$ (=$R_Q$-$G$=$R$-$G$) becomes less.

Because $H(Q||P)$ becomes less after every step, the iteration converges. **Q.E.D.**

Now we use Figure 7 to explain (loosely prove) why the iteration can make $(G, R)$ converge to $(G^*, R^*)$, which should be global convergence instead of local convergence. We need horizontal improvement that makes $(G, R)$ close to line $R=G$, and also longitudinal improvement that makes $R$ close to $R^*$.

**In the Right-step**, $G$ increases, and $R$ remains. So, $(G, R)$ is moved towards rightward so that $(G, R)$ is closer to the line $R=G$ (rather than $G$ is closer to $G^*$), and hence is horizontally improved.

**In the Left-step a**, $Q(X)$ that is used to construct $P(y_j|X)$ becomes $Q^{+1}(X)$. Hence there is the generalized KL information

$$H(Q^{+1}/Q|P) = \sum_i P(x_i)\log\frac{Q^{+1}(x_i)}{Q(x_i)} = \sum_i[\sum_j P(x_i)P*(x_i|y_j)]\log\frac{\sum_j P(y_j)P(x_i|\theta_j)}{\sum_j P(y_j)P^{-1}(x_i|\theta_j)} > 0 \qquad (6.8)$$

To compare it with the average of the generalized KL information:

$$\Delta R_a = \sum_i\sum_j P(x_i)P*(x_i|y_j)\log\frac{P(x_i|\theta_j)}{P^{-1}(x_i|\theta_j)} \qquad (6.9)$$

We can find $\Delta R_a$ and $H(Q^{+1}/Q||P)$ are approximate. So, there is also $\Delta R_a>0$, which means $P(X|\Theta)$ is closer to $P*(X|Y)$ than $P^{-1}(X|\Theta)$, and hence $R$ is improved, or say, $(G, R)$ is longitudinally improved. In this step, $G$ may decrease. $G$ decreases because $G$ approaches $R$ and also approaches $G^*$.

**In the Left-step b**, the iteration for $P^{+1}(Y)$ makes $H(P_Y||P_Y^{+1})=0$. When $R>R^*$, $R_Q$ approaches $R$; when $R<R^*$, $R$ approaches $R_Q$. Hence, $R$ or $R_Q$ approaches $R^*$. In both cases, $(G, R)$ moves to right $R(G)$ function curve, and is closer to line $R=G$. Hence $(G, R)$ is improved in two directions. In the first Left-step b, although $H(Q||P)$ decreases, $G$ exceptionally moves away from $G^*$. This might be because the start-step does not improve $Q(X)$. We need further study for this exception.

In summary, the iteration can make $(G, R)$ globally converges to $(G^*, R^*)$.

## 6.4 Comparing the CM Algorithm with the EM Algorithm

Compared with the other methods for mixture models, the CM algorithm is most similar to the EM algorithm. The comparison of the two algorithms will help us understand the both algorithms better.

### *6.3.1 The difference of two algorithms*

In the Dempaste, Laird, and Rubin's paper [8], which proposes the standard EM algorithm, and Wu's paper [18], which provides the improved convergence proof, the likelihood of a mixture model is expressed as $\log P(X^N|\Theta) \geq L=Q-H$. In terms of the semantic information, there is

$$\log P(X^N|\Theta) = N\sum_i P(x_i)\log P(x_i|\Theta) = N\sum_i P(x_i)\log Q(x_i)$$

$$\geq L = N\sum_i\sum_j P(x_i)P(y_j|x_i)\log\frac{P(x_i,y_j|\theta_j)}{P(y_i|x_i)} \qquad (6.10)$$

$$= N\sum_i\sum_j P(x_i)P(y_i|x_i)\log P(x_i,y_j|\theta_j) - N\sum_i\sum_j P(x_i)P(y_i|x_i)\log P(y_i|x_i)$$

$$= Q - H$$

If we move $P(Y)$ or $P(Y|\Theta)$ from $Q$ into $H$, then $Q$ will become $-NH(X|\Theta)$ and $H$ becomes $-NR_Q$. If we add $NH(X)$ to both sides of the inequality, we will have $H(Q||P) \leq R_Q-G$, which is similar to Eq. (6.7). It is easy to prove

$$Q=NG-NP(X)-NH(Y) \qquad (6.11).$$

where $H(Y)=-\sum_j P^{+1}(y_j)\log P(y_j)$ is a generalized entropy.

The E-step of the EM algorithm and the Left-step a of the CM algorithm (see Eq. (6.1))are the same. The M-step of the EM algorithm maximizes $Q$. It is equal to maximizing $-H(X, Y|\Theta)=G-H(X)-H(Y)$. So, the M-step increases $G$ and decreases $H(Y)$. We may think the M-step merges the Left-step b and the Right-step of the CM algorithm into one step. In summary,

The E-step of EM = the Left-step a of CM

The M-step of EM ≈ the Left-step b + the Right-step of CM

Neal and Hinton proposed a variant EM algorithm [19]. In this algorithm, $F(P_Y,\Theta)=-H(X, Y|\Theta)+H(Y)(H(Y)$ is a Shannon entropy)is used as the objective function for optimization. $F(P_Y,\Theta)$ is similar to $I(X; \Theta)$ in the CM algorithm. In this variant EM algorithm, the E-step also maximize $F(P_Y, \Theta)$. However, in the Left-steps of the CM algorithm, we only need to optimize $P(Y)$, rather than to maximize $G$. For instance, in the Example 2, $G$ is decreasing in all left-steps except the first Light-step b.

There are also other improved EM algorithms [7, 20-23] with some advantages. However, no one of these algorithms facilitates that $R$ converges to $R^*$, and $R-G$ converges to 0 as the CM algorithm.

### 6.3.2 The problem with the convergence proofs of the EM algorithm

The standard EM algorithm in the M-step performs the Left-step b and the Right-step of the CM algorithm at the same time. Since the Left-step b and Right-step have different purposes, it is not easy for the M-step to make $G$ close to $G^*$ or $P(X|Y, \Theta)$ close to $P^*(X|Y)$ in some cases. For instance, in the Example 2, $R^*$ and $G^*$ are very small, and $G$ is decreasing in all Left-steps except for the first Left-step b. Because only when $G$ is decreasing, $G$ can approach less $G^*$. In the Example 1, $H(Y)$ should be increasing because only when $H(Y)$ is increasing, $H(Y)$ can approach larger $H^*(Y)$. However, maximizing $Q$ by the M-step is equivalent to maximizing $G-H(Y)$ always. If we first optimize $P(Y)$ and then optimize $P(X|Y, \Theta)$ in the M-step, then the EM algorithm will be equivalent to the CM algorithm. That is why EM algorithm can converge.

In [8] and [19], the authors prove that $Q$ in the E-step is ne-decreasing. Byrne points out [18] that their proofs for "$Q$ in the E-step is ne-decreasing" are flawed. We argue that "$Q$ in the E-step is ne-decreasing" is not necessary because when $R>R^*$, it impedes $Q$ to converge to $Q^*=-NH^*(X, Y)$. In Example 2, $Q^*=-6.031N$. After the first optimization of parameters, $Q=-6.011N>Q^*$. If we continuously maximize $Q$, $Q$ cannot approach less $G^*$. If $Q$ decreases in E-steps, their convergence proofs cannot be tenable.

A variant EM algorithm proposed by Neal and Hinton [20] moving optimizing $P(Y)$ from the M-step to the E-step to accelerate the convergence. The CM algorithm is similar to this algorithm. Yet, it is questionable to maximize $F(P_Y, \Theta)$ in both steps. For the Example 2, maximizing $G$ or $F(P_Y, \Theta)$ is against making $G$ close to $G^*$.

The Jensen's inequality is used for the EM algorithm so that maximizing $\log P(X^N|\Theta)$ is equivalent to maximizing $L=Q-H$. However, the CM algorithm does not use Jensen's inequality. The reason is that the sampling distribution $P(X)$, sub-models $\theta_j$ ($j=1, 2, …, n$), and the semantic information measures are used so that we have $H(Q||P)=R_Q-G$. Minimizing $H(Q||P)$ is equivalent to maximizing $G$ or $\log P(X^N|\Theta)$.

According to above theoretical analyses and examples, the CM algorithm converges faster and with clearer reasons in comparison with the standard EM algorithm. However, the convergence proof of the CM algorithm still needs improvements so that the proof is stricter in mathematics.

### 6.3.3 Speeds and applications of algorithms

For Gaussian mixture models with $n=2$, we used different true parameters and the same start parameters to examine the convergent speed of the CM algorithm. Assume that when $H(Q||H)\leq 0.001$, the iteration converge. The most possible number of iterations for convergence was 5. In minor cases where $R-G$ was much less than $|G^*-G|$, the iterative numbers for convergence were more than 30. The median number of iterations should be less than 10.

According to [20], the standard EM algorithm needs more than 30 iterations; the variant EM algorithm needs about 17 iterations. According to [19], the standard EM algorithm needs about 18 iterations; the Multi-Set EM algorithm needs about 12 iterations. In [7] where an improved EM algorithm is compared with the Newton method and distance is used as the convergence criterion, the improved EM algorithm needs 17 iterations in average, and the Newton method needs 7 iterations in average. Yet, times that the both algorithms cost are similar.

Obviously, the EM algorithm has higher convergence speed than the standard EM algorithm. To know exact differences of the numbers of iterations and times cost for convergence by two algorithms, we need to compare them with the same true parameters and start parameters. To the minor cases of slow convergence, if we selectively combine those methods of improving the EM algorithm [18, 21-24], such as the methods of optimizing start parameters, with the CM algorithm, the convergence for these cases should be much faster.

The model obtained from the CM algorithm can be used in cases where the source $P(X)$ is changed. According to Eq. (3.12), with $P(X)$ and optimized $P(X|\theta_j) \approx P^*(X|y_j)$ (for all $j$), we can obtain the truth function $T(A_j|X)$, which is similar to the logistic function. When $P(X)$ is changed, we can obtain new likelihood function by the semantic Bayesian formula Eq. (2.4). For example, $X$ denotes stature and $Y$ denotes sex. First, we obtain the optimized likelihood function $P(X|\theta_1)$ with a sample from adult population. Then, we can obtain $T(\theta_1|X)$, and apply it to a specific crowd (such as middle school students) with different $P(X)$ (the mean stature is a little shorter). For given sex (such as $y_1$ for male), we can make semantic Bayesian inference to obtain new likelihood function $P(X|\theta_1)=P(X)T(\theta_1|X)/T(\theta_1)$. However, in popular likelihood methods, the optimized model cannot be used to apply to cases with different $P(X)$.

The CM algorithm can also be used for decision (or classification) function with ML criterion. The Left-step c can be used for this purpose. The EM algorithm does not provide a similar method.

Because the CM algorithm uses sampling distribution instead of sampling sequence, it fits cases with larger samples. However, the EM algorithm fits cases with smaller samples.

The EM algorithm is based on a new semantic information theory. It has many other characteristics, such that it can be more conveniently used for tests and estimations than for mixture models; it may be used to explain the evolution of natural languages as discussed before.

# 7 Conclusions

This paper restates Lu's semantic information method to show that his semantic information is defined with average log normalized likelihood. The paper reveals that by letting the semantic channel and Shannon channel mutually match and iterate (the CM algorithm), we can achieve the maximum Shannon mutual information and maximum average log-likelihood for tests, estimations, and mixture models. The iterative convergence can be intuitively explained and proved by Lu's $R(G)$ function. Several iterative examples of tests, estimations, and mixture models are provided. These examples and theoretical analyses show that, in comparison with the standard EM algorithm, the CM algorithm has faster speed, clearer convergence reasons, and wider potential applications.

The paper also concludes that the tight combination of Shannon information theory and likelihood method is necessary for resolving difficult problems with tests, estimations, and mixture models. The results show that with Lu's semantic information method, the combination is feasible.

**Acknowledgements** The author thanks Professor Peizhuang Wang for his long term supports. The author's earlier monograph *A Generalized Information Theory* was finished when the author studied in Beijing Normal University as a visiting scholar of Professor Wang. Without his recent encouragement, the author wouldn't have worked so hard.